\documentclass[conference]{IEEEtran}
\IEEEoverridecommandlockouts

\usepackage{cite}
\usepackage{amsmath,amssymb,amsfonts}
\usepackage{algorithmic}
\usepackage{graphicx}
\usepackage{textcomp}
\usepackage{xcolor}

\usepackage{hyperref}
\usepackage{tikz}
\DeclareRobustCommand*\circled[1]{%
  \tikz[baseline=(char.base)]{
    \node[shape=circle,draw,inner sep=0.5pt] (char) {#1};
  }%
}

\usepackage[table]{xcolor}
\usepackage{booktabs}
\usepackage{arydshln}
\usepackage{multirow}

\newcommand{\pair}[2]{%
\begin{tabular}[c]{@{}c@{\;\textbar\;}c@{}}
#1 & #2
\end{tabular}}

\newcommand{\perfhead}{%
\multicolumn{2}{c}{} 
& WER$\downarrow$ basic \textbar{} full
& SS$\uparrow$ basic \textbar{} full
& DNSMOS basic \textbar{} full
& MAE$\downarrow$ basic \textbar{} full \\
\cmidrule(lr){3-3}
\cmidrule(lr){4-4}
\cmidrule(lr){5-5}
\cmidrule(lr){6-6}
}

\newcommand{\sotacell}[1]{\textcolor{red}{\textbf{#1}}}
\newcommand{\secondcell}[1]{\textcolor{blue}{\textbf{#1}}}

\def\BibTeX{{\rm B\kern-.05em{\sc i\kern-.025em b}\kern-.08em
    T\kern-.1667em\lower.7ex\hbox{E}\kern-.125emX}}
\begin{document}

\title{CosyEdit: Unlocking End-to-End Speech Editing Capability from Zero-Shot Text-to-Speech Models \\
\thanks{$^{\dag}$ These authors contributed equally to this work.}
\thanks{$^{*}$ Corresponding author.}
}

\author{\IEEEauthorblockN{Junyang Chen$^{1,\dag}$, Yuhang Jia$^{1,\dag}$, Hui Wang$^{1}$, Jiaming Zhou$^{1}$, and Yong Qin$^{1,*}$}
\IEEEauthorblockA{
    $^{1}$College of Computer Science, Nankai University, Tianjin, China\\
    chenjunyang@mail.nankai.edu.cn, qinyong@nankai.edu.cn
    }
}

\maketitle

\begin{abstract}
Automatic speech editing aims to modify spoken content based on textual instructions, yet traditional cascade systems rely on explicit temporal alignment and complex preprocessing. To address these limitations, we propose CosyEdit, an end-to-end speech editing model adapted from CosyVoice through task-specific post-training and a complementary training paradigm, which internalizes text--speech alignment while ensuring high consistency between the speech before and after editing. Trained on only 250 hours of supervised data from our curated GigaEdit dataset, our 400M-parameter model achieves reliable speech editing performance. Extensive evaluations show that CosyEdit not only outperforms several billion-parameter language model baselines but also approaches state-of-the-art cascade systems. These results show that robust and efficient speech editing can be unlocked from a zero-shot TTS model through post-training, offering a cost-effective end-to-end solution for high-quality speech editing. Code and audio samples are available at \url{https://cjy1018.github.io/CosyEditDemoPage/}.
\end{abstract}


\section{Introduction}
Automatic speech editing aims to modify an existing speech recording according to textual instructions, enabling direct insertions, deletions, or substitutions at the audio level without re-recording. Unlike zero-shot text-to-speech (TTS), which primarily focuses on preserving speaker timbre, speech editing further requires preserving the prosodic and paralinguistic consistency of unedited regions while maintaining overall fluency after editing. Achieving natural and reliable edits requires addressing two core challenges: (1) precise cross-modal temporal alignment between speech and text, and (2) context-consistent zero-shot generation for the modified segments.

\begin{figure}
    \centering
    \vspace{-1mm}
    \includegraphics[width=0.96\linewidth]{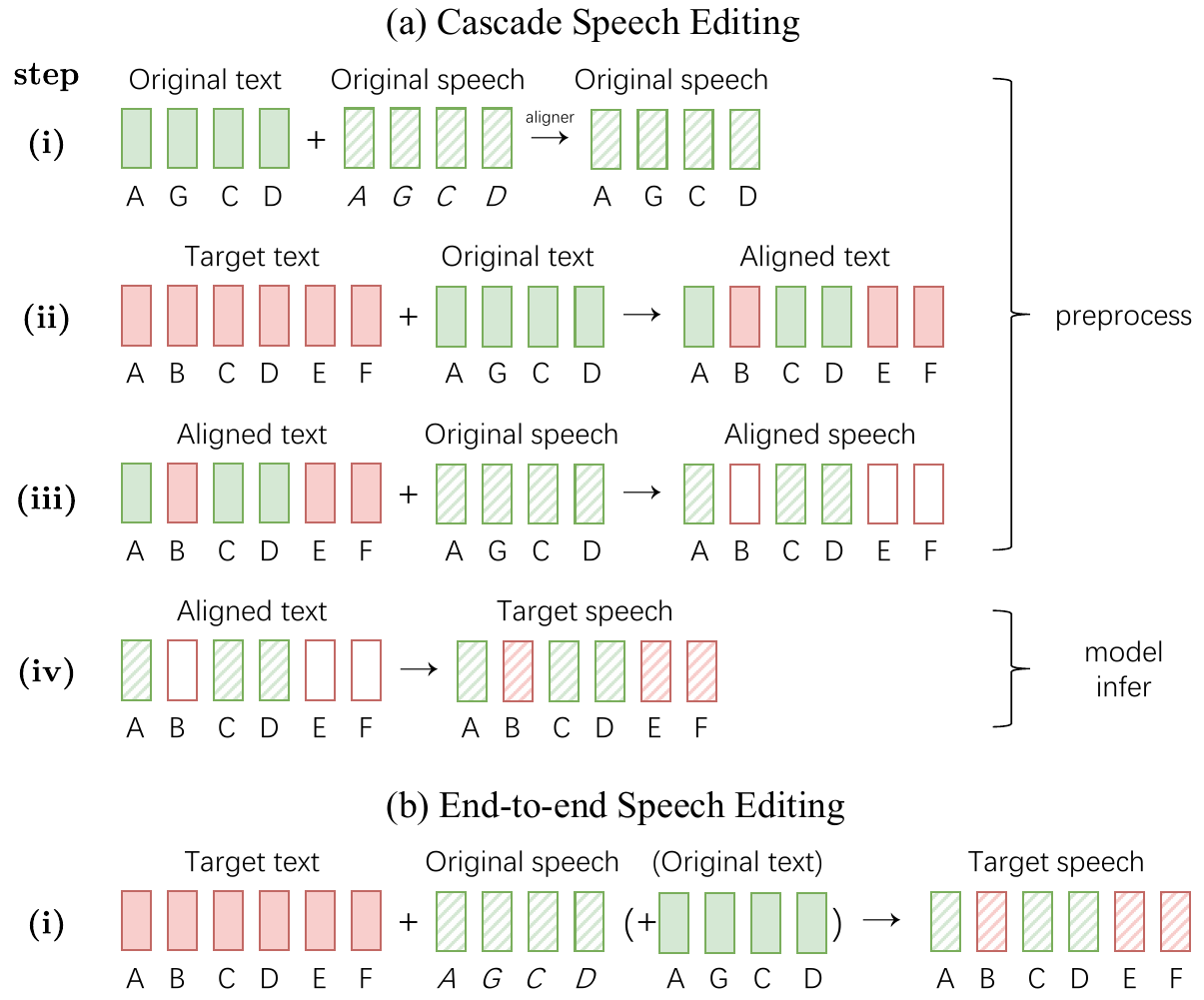}
    \caption{
    Comparison between cascade and end-to-end speech editing. \textit{Italicized} characters indicate speech segments not temporally aligned with the text, while upright characters denote segments with established alignment timestamps. Red empty rectangular boxes represent masked speech tokens to be edited.}
    \label{fig:MFA_and_E2E}
    \vspace{-1mm}
\end{figure}

Early speech editing systems typically rely on external text--speech alignment tools, such as the Montreal Forced Aligner (MFA)~\cite{mcauliffe2017montreal}, to establish temporal correspondence between speech and transcript (Fig.~\ref{fig:MFA_and_E2E}(a), step~(i)). The system then identifies the textual edit span by comparing target and original texts (step~(ii)), determines speech edit boundaries from the aligned text span (step~(iii)), and finally synthesizes the edited segment for integration into the preserved context (step~(iv)). This multi-stage design introduces substantial engineering overhead: each stage requires dedicated tools and careful calibration, and alignment errors in early stages propagate through the pipeline, degrading final output quality.

As summarized in Table~\ref{tab:comparison}, cascade pipelines are typically built on either non-autoregressive (NAR)~\cite{jiang2023fluentspeech, le2023voicebox, chen2025f5, wang2025maskgct} or autoregressive (AR)~\cite{peng2024voicecraft, wang2025ssr} architectures, each with inherent limitations: NAR models require an auxiliary duration predictor~\cite{jiang2023fluentspeech, wang2025maskgct} to avoid prosody mismatches at edit boundaries, while AR models are prone to sampling instability~\cite{peng2024voicecraft} and unnatural boundary transitions~\cite{wang2025ssr} without additional stabilization.

Recent advances in speech language models (SLMs)~\cite{yan2025step, zhang2025mimo, yan2025ming} have introduced a new paradigm for end-to-end speech editing. By internalizing text--speech alignment within a unified representation space, they eliminate the need for external alignment modules and are inherently more amenable to supporting arbitrary edit types and spans within a single inference pass. Despite these advances, the speech editing capability of existing SLMs is largely a byproduct of general-purpose tasks, lacking specialized architectural designs and targeted training objectives tailored for speech editing. Furthermore, these models are typically designed with massive parameter scales and demand extensive large-scale training, prohibitive for academic research and practical deployment.

Motivated by these observations, we explore task-specific post-training as a lightweight way to adapt pretrained zero-shot TTS models for end-to-end speech editing. Our approach is grounded in the insight that speech editing shares core competencies with zero-shot TTS, including: (1) the ability to generate natural speech from text, (2) in-context learning capabilities, and (3) potential for temporal alignment. Based on this strategy, we introduce \textbf{CosyEdit}, a 400M-parameter model initialized from CosyVoice~\cite{du2024cosyvoice} and post-trained with only 250 hours of task-specific supervised editing data. Experimental results demonstrate that CosyEdit achieves state-of-the-art (SOTA) performance in end-to-end speech editing and remains highly competitive against advanced cascaded systems. 

Our contributions are threefold:

\begin{table}
    \centering
    \caption{Comparison of speech editing models.}
    \label{tab:comparison}
    \resizebox{\linewidth}{!}{
        \begin{tabular}{lccccc}
            \toprule
            Method & Arch. & E2E & Multi-Edit & Params & Data \\
            \midrule
            FluentSpeech      & NAR      & $\times$ & $\times$ & 23.9M & 585 h   \\
            VoiceCraft        & AR       & $\times$ & $\times$ & 830M & 10k h   \\
            SSR-Speech        & AR       & $\times$ & $\leqslant 3$ & 830M & 10k h   \\
            \hdashline[2pt/2pt]
            \noalign{\vskip 0.5ex}
            Step-Audio-EditX  & AR+NAR & \checkmark & \checkmark & 3B  & 200k h \\
            MiMo-Audio        & AR+NAR & \checkmark & \checkmark & 7B & 100M h  \\
            Ming-UniAudio     & AR+NAR & \checkmark & $\times$ & 16B & 390k h  \\
            CosyEdit (ours)   & AR+NAR & \checkmark & \checkmark & 400M & 250 h   \\
            \bottomrule
        \end{tabular}
    }
\end{table}
\begin{itemize}
    \item We introduce a general pipeline for constructing supervised speech editing datasets from existing speech corpora and build \textbf{GigaEdit}, a 250-hour editing dataset covering diverse acoustic conditions and editing types.
    \item We extend AR+NAR zero-shot TTS models, exemplified by CosyVoice, with a speech-editing-specific autoregressive token modeling paradigm, reference-guided flow matching, and a complementary mixed in-context training scheme, yielding \textbf{CosyEdit}, an end-to-end speech editing model attainable with only 250 hours of post-training.
    \item Extensive evaluations across multiple speech editing benchmarks demonstrate that lightweight post-training is a practical and effective path for adapting zero-shot TTS foundation models to speech editing.
\end{itemize}

\begin{figure*}[htbp]
\begin{center}
\includegraphics[width=0.98\linewidth]{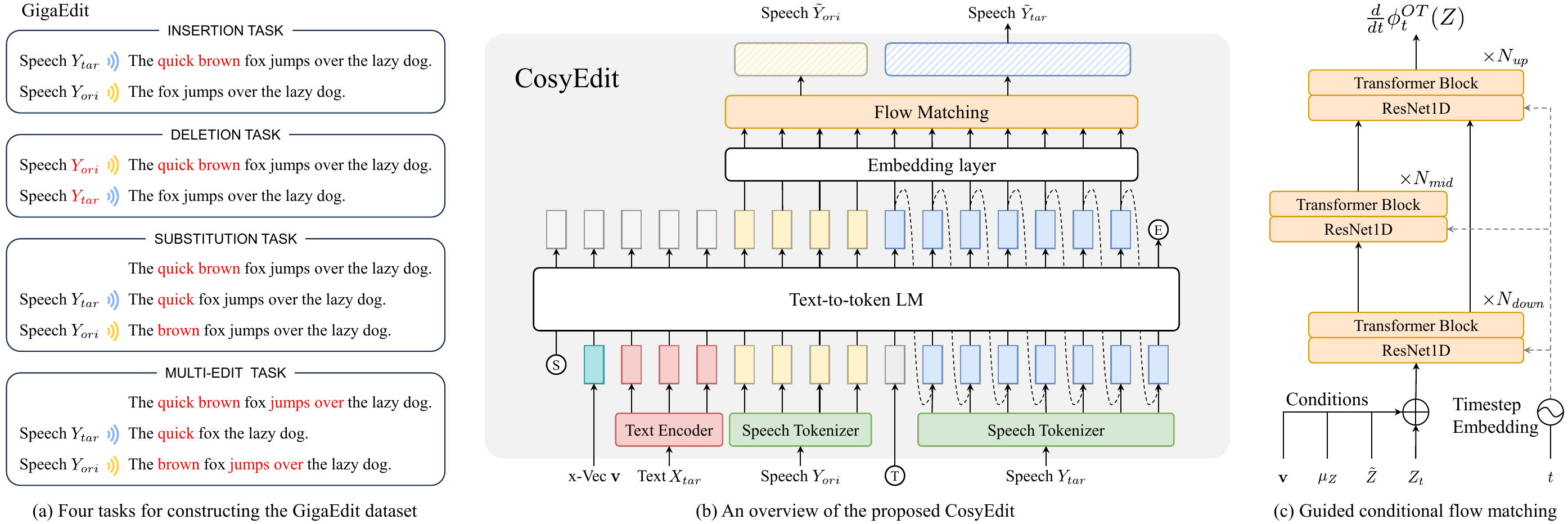}
\end{center}
\vspace{-1mm}
\caption{
(a) Examples of four editing tasks for constructing the GigaEdit training dataset.
(b) Schematic diagram of CosyEdit.
\circled{S}, \circled{E}, and \circled{T} denote the start-of-sequence, end-of-sequence, and transition token, respectively.
Dashed lines indicate the autoregressive decoding.
(c) Enlarged view of the GOT-CFM module, conditioned on speaker embedding $\mathbf{v}$, concatenated semantic tokens $\mu_Z$, concatenated speech features $\tilde{Z}$, and intermediate state $Z_t$ at timestep $t$.
Here, $\mu_Z = [\mu_{\mathrm{ori}}, \mu_{\mathrm{tar}}]$ and $\tilde{Z} = [M_{\mathrm{ori}}, \tilde{M}_{\mathrm{tar}}]$, where $\tilde{M}_{\mathrm{tar}}$ denotes the fully masked target mel-spectrogram.
}
\label{fig:CosyEdit}
\vspace{-1mm}
\end{figure*}

\section{Proposed Method}
Similar to CosyVoice~\cite{du2024cosyvoice}, CosyEdit comprises four components: a text encoder, a $\mathcal{S}^3$ speech tokenizer, an AR large language model (LLM), and a NAR conditional flow-matching (CFM)~\cite{lipman2023flow} model. We retain the original text encoder and $\mathcal{S}^3$ tokenizer while adapting the LLM with task-specific training objectives and redesigning the CFM with reference-guided mechanisms. In addition, we propose two complementary input sequence construction strategies for training to better adapt the model from the rigid, synthetic prosody of TTS systems to real-world, in-the-wild speech editing scenarios.

\subsection{Large Language Model for Speech Editing}
Unlike cascade speech editing approaches that treat editing as masked region prediction conditioned on surrounding context, we reformulate speech editing as an autoregressive discrete speech token generation problem, in which text--speech alignment is implicitly internalized within this process. As illustrated in Fig.~\ref{fig:CosyEdit}(b), we adapt the TTS model to jointly condition on the target text and the original speech, where the original speech provides contextual and acoustic information, while the target text specifies the desired edits. Accordingly, we design the LLM to model the following sequence:
\begin{equation}
    \left[\circled{S},\, \mathbf{v},\, \overline{X}_{tar},\, \mu_{\mathrm{ori}},\, \circled{T},\, \mu_{\mathrm{tar}},\, \circled{E} \right],
\end{equation}
where $\circled{S}$ and $\circled{E}$ denote start and end tokens. The vector \(\mathbf{v}\) is a speaker embedding extracted from the original speech \(Y_{ori}\) using a pretrained speaker-verification model. The text encoding \(\overline{X}_{tar} = \{\bar{x}_u\}_{u\in[1:U]}\) is produced using the same byte-pair encoding (BPE) tokenizer and text encoder as CosyVoice:
\begin{equation}
    \overline{X}_{tar} = \text{TextEncoder}(\text{BPE}(X_{tar})).
\end{equation}

We use the $\mathcal{S}^3$ tokenizer to extract discrete supervised semantic tokens from the original speech and the target speech:
\begin{equation}
\begin{aligned}
    \mu_{\mathrm{ori}} &= \text{SpeechTokenizer}(Y_{ori}),\\
    \mu_{\mathrm{tar}} &= \text{SpeechTokenizer}(Y_{tar}).
\end{aligned}
\end{equation}

Then we insert a single transition token \(\circled{T}\) between the original speech-token sequence \(\mu_{\mathrm{ori}}\) and the target speech-token sequence \(\mu_{\mathrm{tar}}\) to mark the boundary between conditioning context and autoregressive generation. The training objective for the AR token language model is:
\begin{equation}
\mathcal{L}_{LM}
=
-\frac{1}{N+1}\sum_{j=1}^{N+1}\log q(\mu_{\mathrm{tar},j}),
\end{equation}
where $N = |\mu_{\mathrm{tar}}|$, $\mu_{\mathrm{tar},N+1} = \circled{E}$ is the end token, and $q(\mu_{\mathrm{tar},j})$ denotes the predicted probability of the $j$-th target semantic token conditioned on the input sequence and the ground-truth prefix $\mu_{\mathrm{tar},1:j-1}$ under teacher forcing.

\subsection{Guided Optimal-Transport Conditional Flow Matching}
The flow model in CosyVoice is originally designed for zero-shot TTS, where the primary objective is to synthesize clean, studio-quality speech with globally consistent speaker timbre. In contrast, speech editing requires maintaining the complex acoustic environment of real-world recordings beyond speaker identity alone. To this end, we introduce Guided OT-CFM (GOT-CFM), a reference-guided variant of OT-CFM~\cite{tong2023improving}. Specifically, we construct the flow-matching path over the temporal concatenation of original and target mel-spectrograms, where the observable trajectory from noisy to clean original mel-spectrogram serves as an explicit acoustic reference to guide the flow modeling of the target mel-spectrogram. This design allows the flow-matching module to access the full speech context. The training objective is:
\begin{equation}
\begin{aligned}
\mathcal{L}_{GOT\text{-}CFM}
&= \mathbb{E}_{t,\,p_0(Z_0),\,q(Z_1)}
\Big|\, \omega_t\big(\phi_t^{OT}(Z_0,Z_1)\mid Z_1\big) \\
&\quad - \nu_t\big(\phi_t^{OT}(Z_0,Z_1)\mid\theta\big) \Big|,
\end{aligned}
\end{equation}
where
\begin{equation}
Z_0 = [M_{\mathrm{ori}}^{0},\; M_{\mathrm{tar}}^{0}],
\qquad
Z_1 = [M_{\mathrm{ori}},\; M_{\mathrm{tar}}].
\end{equation}

Here, $M_{\mathrm{ori}}^{0}$ and $M_{\mathrm{ori}}$ denote the noisy and clean mel-spectrograms of the original speech, and $M_{\mathrm{tar}}^{0}$ and $M_{\mathrm{tar}}$ denote those of the target speech. The operator $[\cdot,\cdot]$ denotes concatenation along the temporal dimension. The interpolation path $\phi_t^{\text{OT}}(Z_0, Z_1)$ linearly blends $Z_0$ and $Z_1$ over time, while the target vector field $\omega_t(\phi_t^{\text{OT}}(Z_0, Z_1)\mid Z_1)$ provides a constant direction from the noisy state toward the target.

To construct the guiding probability density path, we condition the model on the fully revealed original mel-spectrogram $M_{\mathrm{ori}}$ and the fully masked target mel-spectrogram $\tilde{M}_{\mathrm{tar}}$. The known trajectory from $M_{\mathrm{ori}}^{0}$ to $M_{\mathrm{ori}}$ serves as a guide, encouraging $M_{\mathrm{tar}}^{0}$ to follow a similar path toward $M_{\mathrm{tar}}$. The speaker embedding $\mathbf{v}$, the speech tokens $\mu_Z$, and the concatenation of $M_{\mathrm{ori}}$ and $\tilde{M}_{\mathrm{tar}}$ are fed into the neural network to match the vector field parameterized by $\theta$:
\begin{equation}
\begin{aligned}
&\nu_{t}\!\left(\phi_{t}^{OT}\!\left(Z_{0}, Z_{1}\right) \mid \theta\right) \\
&= \mathrm{NN}_{\theta}\!\Big(\phi_{t}^{OT}\!\left(Z_{0}, Z_{1}\right), t;\;
    \mathbf{v},\; \mu_Z,\; [M_{\mathrm{ori}},\, \tilde{M}_{\mathrm{tar}}] \Big),
\end{aligned}
\end{equation}
where
\begin{equation}
\mu_{Z} = [\mu_{\mathrm{ori}},\; \mu_{\mathrm{tar}}].
\end{equation}

\subsection{Zero-shot and One-shot In-Context Learning}
\label{zicl_and_oicl}
\begin{figure}
    \centering
    \vspace{-1mm}
    \includegraphics[width=0.94\linewidth]{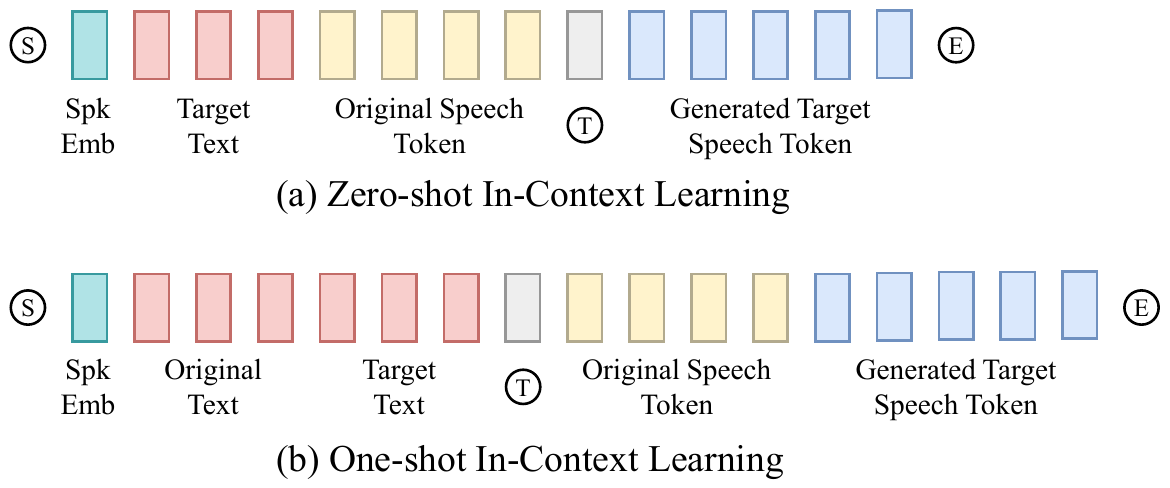}
    \caption{Training sequence format of ZICL and OICL.}
    \label{fig:zicl_and_oicl}
    \vspace{-2mm}
\end{figure}

To improve the model's ability to internalize text--speech alignment while remaining sensitive to localized editing instructions, we introduce two complementary training paradigms, namely \textbf{one-shot in-context learning (OICL)} and \textbf{zero-shot in-context learning (ZICL)}. As illustrated in Fig.~\ref{fig:zicl_and_oicl}, both paradigms share the same autoregressive target speech prediction objective, differing only in whether an explicit original text--speech pair is provided as alignment context.

Given the original text encoding $\overline{X}_{ori}$, target text encoding $\overline{X}_{tar}$, original speech token $\mu_{\mathrm{ori}}$, and target speech token sequence $ \mu_{\mathrm{tar}}$, the input sequence of OICL is constructed as
\begin{equation}
\mathcal{S}_{\mathrm{OICL}}
=
[\circled{S}, \mathbf{v}, \overline{X}_{ori}, \overline{X}_{tar}, \circled{T}, \mu_{\mathrm{ori}}],
\end{equation}
where $\circled{T}$ denotes the transition token separating textual and acoustic contexts in OICL. Conditioned on $\mathcal{S}_{\mathrm{OICL}}$, the model is trained to autoregressively predict $\mu_{\mathrm{tar}}$.

In contrast, ZICL deliberately removes the original text encoding $\overline{X}_{ori}$ from the conditioning context and retains the target text encoding along with the original speech token:
\begin{equation}
\mathcal{S}_{\mathrm{ZICL}}
=
[\circled{S}, \mathbf{v}, \overline{X}_{tar}, \mu_{\mathrm{ori}}, \circled{T}],
\end{equation}
where the model predicts the same target speech token $ \mu_{\mathrm{tar}}$. To avoid introducing inconsistent text--speech associations across training paradigms, $\mu_{\mathrm{ori}}$ is placed before $\circled{T}$ in ZICL, ensuring that the original speech is consistently treated as conditioning context rather than an autoregressive target prefix.

The two paradigms present complementary trade-offs. OICL supplies paired original text and speech, providing accurate implicit temporal alignment references. This explicit supervision, however, risks inducing shortcut behavior, where the model over-relies on alignment cues and tends to favor copying the original speech rather than performing the intended edits. ZICL, by contrast, withholds the original text entirely, compelling the model to infer edit boundaries from the discrepancy between the target text and the original speech, thereby strengthening instruction-following at edited positions, albeit at the cost of losing direct alignment supervision.

To balance retention fidelity against editing capability, ZICL and OICL samples are mixed during training at a ZICL mixing ratio $\lambda$, yielding the composite objective
\begin{equation}
\mathcal{L}
=
\lambda \mathcal{L}_{\mathrm{ZICL}}
+
(1-\lambda)\mathcal{L}_{\mathrm{OICL}},
\end{equation}
where
\begin{equation}
\mathcal{L}_{*}
=
-\log P(\mu_{\mathrm{tar}}|\mathcal{S}_{*}).
\end{equation}

\section{Experiments}
\subsection{Experimental Setup}
\noindent\textbf{GigaEdit Dataset.} We propose a data construction procedure that is able to transform existing speech corpora into supervised speech editing datasets covering insertion, deletion, and substitution sub-tasks. Using this procedure, we construct the GigaEdit dataset based on GigaSpeech-S~\cite{chen2021gigaspeech}. As illustrated in Fig.~\ref{fig:CosyEdit}(a), we treat each utterance and its transcript as the target speech and target text, and use MFA to obtain their time alignment. We then construct data for three basic editing types: (1) For the insertion sub-task, we randomly remove some segments of the target speech according to the time alignment (removed length $< |X_{\mathrm{tar}}| / 2$), and the resulting shortened speech and transcript serve as the original speech and original text. (2) The deletion sub-task can be regarded as the symmetric counterpart of the insertion task: we apply the same procedure as for insertion but swap the roles of the original and the target. (3) For the substitution sub-task, we delete a contiguous segment from the target speech, split this segment into two non-overlapping parts, and respectively insert each part back into the deletion site to form two utterances, which are therefore used as a substitution pair.

To improve generalization to scenarios involving multiple edit locations and diverse edit operations, we extend the substitution procedure to a multi-edit task. In this variant, we randomly delete multiple non-contiguous segments from the target speech, while keeping the remaining steps identical to those of the substitution sub-task. The corresponding transcript pairs are generated using the same procedures, enabling the simulation of hybrid basic editing types.

\noindent\textbf{Training Details.} We trained CosyEdit on the GigaEdit dataset using two NVIDIA A800 GPUs. Both the LLM and the flow model were trained for 16 epochs, with learning rates of \(3\times10^{-6}\) and \(1\times10^{-4}\), respectively. During inference, we use the OICL paradigm for better context utilization.

\subsection{Experiments on RealEdit}
We evaluate CosyEdit on RealEdit~\cite{peng2024voicecraft}, a challenging in-the-wild speech editing benchmark comprising 310 samples with diverse and complex acoustic conditions.

\noindent\textbf{Baselines.} Both cascade and end-to-end speech editing systems are included as baselines. The cascade baselines include the AR models VoiceCraft~\cite{peng2024voicecraft}, SSR-Speech~\cite{wang2025ssr}, and the NAR model FluentSpeech~\cite{jiang2023fluentspeech}. The end-to-end baselines comprise Step-Audio-EditX~\cite{yan2025step}, MiMo-Audio~\cite{zhang2025mimo}, and Ming-UniAudio~\cite{yan2025ming}. Although Step-Audio-EditX and MiMo-Audio are not specifically designed for speech content editing, their large-scale training confers a certain degree of generalization to content editing tasks, making them relevant end-to-end baselines. As end-to-end models typically regenerate the entire utterance, we pay particular attention to whether modifications introduced in unedited regions are perceptually noticeable.

FluentSpeech uses the LibriTTS~\cite{zen2019libritts} trained checkpoint with sequential editing for multi-span cases. VoiceCraft follows the silence-reduction strategy of generating five outputs and selecting the shortest. Step-Audio-EditX is run in clone mode with zero-shot inference. MiMo-Audio is run in dialogue mode using five high-quality editing examples generated by SSR-Speech on RealEdit~\cite{peng2024voicecraft} as few-shot prefix prompts, and allows up to five inference attempts to obtain an output whose transcription matches the target text. Ming-UniAudio converts edit prompts into natural-language instructions via a rule-based mapping and applies sequential editing for multi-span cases.

\noindent\textbf{Metrics.} Objective metrics include word error rate (WER, \%) and speaker similarity (SpkSIM, \%), computed using Whisper-medium.en\footnote{\href{https://huggingface.co/openai/whisper-medium.en}{https://huggingface.co/openai/whisper-medium.en}}~\cite{radford2023robust} and WavLM-TDCNN\footnote{\href{https://huggingface.co/microsoft/wavlm-base-plus-sv}{https://huggingface.co/microsoft/wavlm-base-plus-sv}}~\cite{chen2022wavlm}, respectively. Perceptual quality is estimated using two neural MOS predictors, MOSNet~\cite{cooper2022generalization} and UTMOS~\cite{saeki2022utmos}. We also report the mean absolute error MOS between generated and original speech, denoted as ${\text{MAE}_{\text{MOSNet}}}$ and ${\text{MAE}_{\text{UTMOS}}}$. For end-to-end models, we evaluate unedited-region consistency using mel-cepstral distortion (MCD~\cite{kubichek1993mel}; lower is better), computed via dynamic time warping (DTW)~\cite{sakoe1978dynamic} with pymcd\footnote{\href{https://github.com/chenqi008/pymcd}{https://github.com/chenqi008/pymcd}} to compensate for minor temporal misalignments introduced by the forced aligner when extracting unedited regions.

\begin{figure}
    \centering
    \includegraphics[width=0.81\linewidth]{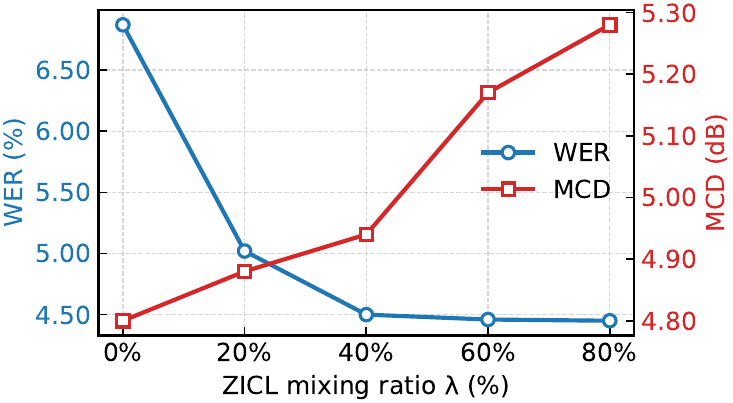}
    \caption{WER and MCD on RealEdit under different ZICL ratios $\lambda$.}
    \label{fig:zicl_ratio}
\end{figure}

\begin{table*}[th]
    \centering
    \caption{Results for speech editing on RealEdit. Bold numbers indicate the best result within each model group separated by the dashed line. * indicates ratings based on speech intelligibility only.}
    \label{tab:main_results}
    \begin{tabular}{c c c c >{\color{black!75}}c c >{\color{black!75}}c c c c}
        \toprule
        \multirow{2}{*}{\raisebox{-0.8ex}{Method}}
        & \multicolumn{7}{c}{Objective Evaluation} 
        & \multicolumn{2}{c}{Subjective Evaluation} \\
        \cmidrule(lr){2-8} \cmidrule(lr){9-10}
        
        & WER $\downarrow$
        & SpkSIM $\uparrow$
        & MCD $\downarrow$
        & MOSNet
        & $\text{MAE}_{\text{MOSNet}} \downarrow$
        & UTMOS
        & $\text{MAE}_{\text{UTMOS}} \downarrow$
        & EMOS $\uparrow$
        & SMOS $\uparrow$ \\
        \midrule

        Ground Truth       
        & 6.06  & -- & -- & 3.34 & --   & 3.38 & --  & 4.21* & -- \\
        \midrule
        
        FluentSpeech     
        & 5.97  & 92.74 & -- & 2.72 & 0.78 & 2.81 & 0.67 & 2.7 & 2.6 \\
        VoiceCraft       
        & 6.55  & 97.12 & -- & 3.18 & 0.24 & 3.31 & 0.20 & 4.04 & 4.08 \\
        SSR-Speech       
        & \textbf{5.05}  & \textbf{98.31} & --
        & 3.32 & \textbf{0.14} & 3.34 & \textbf{0.12} & \textbf{4.11} & \textbf{4.09} \\

        \hdashline[2pt/2pt] \noalign{\vskip 0.5ex}

        Step-Audio-EditX 
        & 10.76 & 95.88 & 8.64 & 3.94 & 0.61 & 3.89 & 0.54 & 3.41 & 3.49 \\
        MiMo-Audio       
        & 16.86 & 93.71 & 9.78 & 3.48 & 0.50 & 3.38 & 0.47 & 3.55 & 3.05 \\
        Ming-UniAudio    
        & 9.98  & 96.70 & 5.36 & 3.13 & 0.33 & 3.18 & 0.30 & 3.79 & 3.84 \\

        \rowcolor{gray!15}
        CosyEdit (ours)             
        & \textbf{4.50}  & \textbf{97.34} & \textbf{4.94}
        & 3.19 & \textbf{0.29} & 3.30 & \textbf{0.25} & \textbf{4.15} & \textbf{4.04} \\

        \bottomrule
    \end{tabular}
\end{table*}
\begin{table}[t]
    \centering
    \setlength{\tabcolsep}{3.2pt}
    \caption{Ablation study on components.}
    \label{tab:ablation_study}
    \resizebox{\linewidth}{!}{
        \begin{tabular}{lccccc}
            \toprule
            Method
            & WER $\downarrow$
            & SpkSIM $\uparrow$
            & MCD $\downarrow$
            & $\text{MAE}_{\text{MOSNet}}\downarrow$ 
            & $\text{MAE}_{\text{UTMOS}}\downarrow$ \\
            \midrule
    
            CosyVoice
            & 4.49 & 95.90 & 6.82 & 0.63 & 0.49 \\
            \quad w/ LLM SFT
            & 5.33 & 96.63 & 6.17 & 0.57 & 0.45 \\
            \quad w/ Flow SFT
            & \textbf{4.18} & 96.73 & 5.59 & 0.31 & 0.27 \\
            
            \midrule
            
            CosyEdit
            & 4.50 & \textbf{97.34} & 4.94
            & \textbf{0.29} & \textbf{0.25} \\
    
            \bottomrule
        \end{tabular}
    }
\end{table}

For subjective evaluation, we randomly sample 10 examples per editing task in RealEdit, including insertion, deletion, substitution, and mixed-edit, yielding 40 samples in total, and collect human ratings for all systems. We introduce two speech-editing-specific metrics beyond conventional MOS: Edit MOS (EMOS) emphasizes semantic aspects, including edit correctness, speech intelligibility and boundary naturalness, whereas Similarity MOS (SMOS) focuses on acoustic consistency, assessing timbre similarity, prosodic appropriateness in edited regions, and preservation of unedited regions. Ten listeners rate each sample on a five-point Likert scale.

\noindent\textbf{Effect of ZICL Mixing Ratio.} Fig.~\ref{fig:zicl_ratio} shows the impact of the ZICL mixing ratio $\lambda$. Increasing $\lambda$ consistently improves editing accuracy, reducing WER from 6.87\% to 4.45\%, while gradually degrading unedited-region fidelity, as reflected by the increase in MCD from 4.80 dB to 5.28 dB. This trend validates the complementary roles of OICL and ZICL discussed in Sec.~\ref{zicl_and_oicl}. We choose $\lambda=0.4$ for all subsequent experiments, as it provides a favorable trade-off between instruction following and preservation fidelity.

\noindent\textbf{Experimental Results.} Table~\ref{tab:main_results} compares cascade speech editing pipelines and end-to-end models on the RealEdit benchmark. CosyEdit surpasses all baselines on both WER and EMOS metrics, demonstrating its strong capability in synthesizing accurate and robust content edits across complex acoustic environments. In terms of acoustic consistency relative to the ground truth (the original speech), as reflected by SpkSIM and SMOS metrics, CosyEdit outperforms all end-to-end baselines and exceeds several traditional cascade systems, approaching the performance of SSR-Speech. For perceptual quality, measured by $\text{MAE}_{\text{MOSNet}}$ and $\text{MAE}_{\text{UTMOS}}$, CosyEdit achieves the lowest overall quality deviation before and after editing among end-to-end models, indicating that the edited speech maintains synthesis quality close to the original speech.

\noindent\textbf{Ablation Study.} We investigate the effect of speech-editing-specific fine-tuning on the LLM and Flow modules, with results shown in Table~\ref{tab:ablation_study}. Using the zero-shot TTS CosyVoice as the baseline, adding LLM SFT slightly increases WER from 4.49 to 5.33. Detailed analysis indicates that this is mainly due to substitution errors where Whisper confuses phonetically similar words. Importantly, LLM SFT improves prosodic consistency in unedited regions, making the rhythm more consistent with the in-the-wild original speech, rather than the studio-level rhythm of the zero-shot TTS. This results in better acoustic consistency, as reflected by reductions in MCD and MAE MOS. SpkSIM also shows a modest increase, suggesting an improvement in speaker timbre preservation. 

In contrast, Flow SFT on GOT-CFM yields consistent improvements across all objective metrics. Subjective inspection indicates that Flow fine-tuning primarily enhances acoustic detail modeling, leading to clearer distinctions between phonetically similar words and acoustic realizations that more faithfully match the original speech. The gains from LLM SFT and Flow SFT are complementary and largely additive. Combining both components results in the final CosyEdit system, which achieves the best overall balance between editing accuracy, acoustic consistency, and speaker preservation.

\begin{figure}
    \centering
    \includegraphics[width=0.98\linewidth]{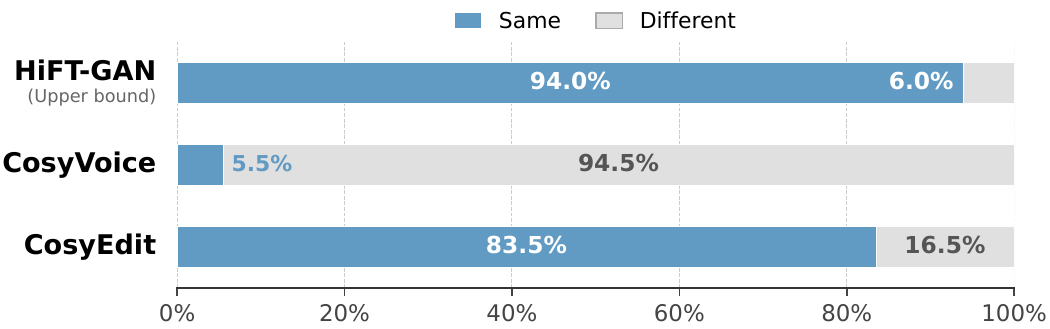}
    \caption{Perceptual transparency test. Listeners judged whether reconstructed speech was perceptually the same as the original.}
    \label{fig:model_comparison}
\end{figure}

\noindent\textbf{Perceptual Transparency Test.} To assess perceptual changes in unedited regions, we conducted a perceptual transparency test in which listeners made binary same/different judgments between the original and reconstructed speech. We randomly selected 40 utterances from RealEdit, using CosyEdit and the zero-shot TTS mode of CosyVoice to synthesize target speech with the same semantics as the original speech. Additionally, we used CosyVoice's HiFT-GAN vocoder~\cite{kong2020hifi, li2023hiftnet} as an upper bound by reconstructing waveforms from the original speech's mel spectrograms. Five listeners evaluated whether they could perceive any differences between the reconstructed and original speech, including content errors, speaker variations, prosody changes, audio quality, and background artifacts, using a binary (same/different) scoring scheme.

As shown in Fig.~\ref{fig:model_comparison}, CosyEdit achieves 83.5\% "same" ratings, substantially higher than CosyVoice zero-shot TTS (5.5\%), and approaching the HiFT-GAN upper bound (94.0\%), demonstrating a clear improvement over zero-shot TTS in maintaining perceptual similarity to the original speech.

\begin{table*}[t]
\centering
\renewcommand{\arraystretch}{1.1}
\caption{Performance comparison on the English subset of Ming-Freeform-Audio-Edit. MAE denotes \(\mathrm{MAE}_{\mathrm{DNSMOS}}\) between generated and original speech, where lower is better. \sotacell{red} and \secondcell{blue} denote the best and second-best system results, respectively.}
\label{tab:editing_results_en_dnsmos_mae_split}
\begin{tabular}{llcccc}
\toprule
Edit Type & Model & \multicolumn{4}{c}{Performance} \\
\midrule
\perfhead
\multirow{5}{*}{Insertion}
& Ground Truth   & \pair{--}{--} & \pair{--}{--} & \pair{2.99}{3.01} & \pair{--}{--} \\
& VoiceCraft     & \pair{4.39}{4.51} & \pair{0.85}{0.85} & \pair{3.01}{3.03} & \pair{\secondcell{0.146}}{\secondcell{0.144}} \\
& SSR-Speech     & \pair{\sotacell{1.75}}{\sotacell{2.03}} & \pair{\sotacell{0.94}}{\sotacell{0.94}} & \pair{3.06}{3.06} & \pair{\sotacell{0.139}}{\sotacell{0.128}} \\
& Ming-UniAudio  & \pair{6.49}{7.84} & \pair{0.80}{0.80} & \pair{3.04}{3.04} & \pair{0.168}{0.167} \\
& CosyEdit       & \pair{\secondcell{2.83}}{\secondcell{2.85}} & \pair{\secondcell{0.86}}{\secondcell{0.86}} & \pair{3.10}{3.11} & \pair{0.167}{0.167} \\
\midrule
\perfhead
\multirow{5}{*}{Deletion}
& Ground Truth   & \pair{--}{--} & \pair{--}{--} & \pair{3.05}{3.03} & \pair{--}{--} \\
& VoiceCraft     & \pair{6.07}{\secondcell{5.86}} & \pair{0.81}{0.81} & \pair{3.01}{3.00} & \pair{0.166}{0.168} \\
& SSR-Speech     & \pair{\sotacell{5.22}}{\sotacell{5.29}} & \pair{\sotacell{0.91}}{\sotacell{0.91}} & \pair{3.03}{3.02} & \pair{\sotacell{0.132}}{\sotacell{0.134}} \\
& Ming-UniAudio  & \pair{14.79}{24.37} & \pair{0.77}{0.75} & \pair{2.97}{2.97} & \pair{0.206}{0.204} \\
& CosyEdit       & \pair{\secondcell{5.69}}{5.95} & \pair{\secondcell{0.83}}{\secondcell{0.83}} & \pair{3.10}{3.09} & \pair{\secondcell{0.161}}{\secondcell{0.164}} \\
\midrule
\perfhead
\multirow{5}{*}{Substitution}
& Ground Truth   & \pair{--}{--} & \pair{--}{--} & \pair{3.04}{3.05} & \pair{--}{--} \\
& VoiceCraft     & \pair{3.13}{2.96} & \pair{0.80}{0.81} & \pair{3.02}{3.02} & \pair{0.172}{0.164} \\
& SSR-Speech     & \pair{\sotacell{1.90}}{\sotacell{1.95}} & \pair{\sotacell{0.89}}{\sotacell{0.90}} & \pair{3.08}{3.08} & \pair{\sotacell{0.146}}{\sotacell{0.140}} \\
& Ming-UniAudio  & \pair{8.10}{7.95} & \pair{0.77}{0.77} & \pair{3.00}{3.03} & \pair{0.166}{0.178} \\
& CosyEdit       & \pair{\secondcell{2.61}}{\secondcell{2.56}} & \pair{\secondcell{0.83}}{\secondcell{0.83}} & \pair{3.11}{3.13} & \pair{\secondcell{0.151}}{\secondcell{0.146}} \\
\bottomrule
\end{tabular}
\end{table*}

\subsection{Additional Experiments on Ming-Freeform-Audio-Edit}
We further evaluate CosyEdit on the English semantic-editing subset of the Ming-Freeform-Audio-Edit~\cite{yan2025ming} benchmark. Compared with RealEdit, this dataset explicitly categorizes editing operations into insertion, deletion, and substitution, allowing for separate analysis of each type. We report results on both the basic and full subsets.

\noindent\textbf{Baselines.} We compare CosyEdit against representative speech editing systems, including VoiceCraft~\cite{peng2024voicecraft}, SSR-Speech~\cite{wang2025ssr}, and Ming-UniAudio~\cite{yan2025ming}. All baselines are evaluated under the same settings used in the RealEdit benchmark.

\noindent\textbf{Metrics.}
Following the benchmark protocol, we report WER for editing correctness, speaker similarity (SS) for speaker preservation, and DNSMOS~\cite{reddy2022dnsmos} for perceptual speech quality. Importantly, unlike standard TTS evaluation where higher DNSMOS is directly preferred, speech editing requires the generated utterance to preserve the acoustic quality of the original recording rather than simply maximize predicted quality. We therefore additionally report $\mathrm{MAE}_{\mathrm{DNSMOS}}$ to measure the mean absolute deviation of predicted MOS between generated and original speech, where lower values indicate better quality consistency before and after editing.

\noindent\textbf{Experimental Results.} 
Table~\ref{tab:editing_results_en_dnsmos_mae_split} reports performance on Ming-Freeform-Audio-Edit. CosyEdit exhibits strong and stable performance across all three edit types. On editing correctness, it achieves the second-best WER for insertion (2.83/2.85) and substitution (2.61/2.56), and remains competitive on deletion (5.69/5.95), substantially outperforming the large end-to-end baseline Ming-UniAudio, which highlights the effectiveness of task-specific post-training for content editing over general-purpose speech language modeling.

On the acoustic side, CosyEdit consistently ranks second in speaker similarity (SS), closely approaching SSR-Speech across all edit types, indicating robust preservation of speaker identity. $\mathrm{MAE}_{\mathrm{DNSMOS}}$ further confirms that CosyEdit maintains a perceptual quality closer to the original speech than end-to-end baseline, exhibiting smaller deviations.

While SSR-Speech achieves the best performance, its cascade system with explicit text--speech alignment and localized reconstruction provides a favorable inductive bias for content-preserving edits. In contrast, CosyEdit operates fully end-to-end without requiring external alignment tools or manually specified edit boundaries at inference time. Overall, these results underscore a favorable trade-off between accuracy and simplicity. Although CosyEdit does not surpass the strongest cascade system, it delivers competitive performance within a single model architecture. This lightweight post-training framework significantly reduces training cost, enhances deployment ease, and ultimately achieves well-balanced gains across WER, speaker similarity, and acoustic consistency.

\section{Conclusions}
In this paper, we propose CosyEdit, an end-to-end speech editing model that internalizes temporal alignment, eliminating external modules and complex preprocessing at inference time. Instead of training large-scale speech language models from scratch, we introduce a task-specific post-training framework for AR+NAR zero-shot TTS models, enabling efficient and cost-effective adaptation for speech editing. Fine-tuned on 250 hours of the GigaEdit dataset, CosyEdit outperforms recent end-to-end baselines and approaches SOTA cascade systems. To support responsible use, we further highlight the importance of mitigating potential misuse in speech deepfakes. To this end, we release the code and datasets to facilitate research on watermarking and partially edited audio detection. Future work will focus on multilingual extension, finer-grained control, and minimizing distortion in unedited regions.

\bibliographystyle{IEEEtran}
\bibliography{mybib}

\end{document}